# Rasch Analysis of the Mathematics Self Concept Questionnaire

Jiaqi Cai

Boston College

## Instrument and Sample

  TIMSS & PIRLS International Study Center is a research center at Boston College that conducts a series of assessments in a number of countries to measure trends in mathematics and science achievement at the fourth and eighth grades. In general, TIMSS assessments include achievement tests as well as questionnaires for student, parent, teacher, school, and curricular. There are 63 participating countries and 14 benchmarking participants in TIMSS 2011, including 608,641 students, 49,429 teachers, 19,612 school principals, and the National Research Coordinators of each country. TIMSS data are valuable for researchers and analysts from all over the world, especially for those from participating countries, to conduct related studies to improve education.

  TIMSS believes that context can directly influence students' learning and educational achievement. Therefore, several questionnaires are conducted and the results are able to help people better understand the achievement results as well as improve education more effectively. TIMSS 2011 conducts five questionnaires for fourth grade; they are Fourth Grade Student Questionnaire, Fourth Grade Home Questionnaire, Fourth Grade Teacher Questionnaire, Fourth Grade School Questionnaire, and Fourth Grade Curriculum Questionnaire. Questions are developed by collaboration between TIMSS & PIRLS, country representatives, and policy experts.

  68 questions are included in TIMSS 2011 student questionnaire; they can be categorized into four groups, background about student, background about student's school, Mathematics in school, and Science in school. There is no time limit to complete the survey. Before answering questions, student may need to spend a few minutes to take a look at 3 examples listed on the top of survey in order to be familiar with types of survey questions. The majority of questions have four options, but some are binary or multiple-choice questions appear in the section of background about student.

This article analyzes 7 items extracted from TIMSS 2011 student questionnaire for Taiwan, focusing on students' attitudes toward their mathematics ability. The dataset includes data from 4027 fourth graders. The data of Taiwan were chosen because the mathematical education in Taiwan is similar to China that I'm most familiar with. Intensity and difficulty are two characteristics of mathematical education in Taiwan. The school and society pay a lot of attention on math education in elementary school, requiring students to acquire skills to solve more difficult mathematical problems comparing with the mathematical requirements in most of other countries in the world. Since students receive more mathematical trainings, students in Taiwan are expected to have higher performance in math achievement tests and feel more confident about their mathematical ability.

In general, scoring rule in this analysis is: agree a lot = 4 point, agree a little = 3 points, disagree a little = 2 points, disagree a lot = 1 points. For three items (03B, 03C, 03G) that are directional inversed, scoring rule is reversed: agree a lot = 1 point, agree a little = 2 points, disagree a little = 3 points, disagree a lot = 4 points.

In this article, the variable that is proposed test is students' mathematics self-concept. According to PISA2012, "students' mathematics self-concept, or belief in their own abilities, is an important outcome of education and strongly related to successful learning". These 7 items measures students' mathematics self-concept by using students' responses as to whether they strongly agree, agree, disagree or strongly disagree with statements of items.

In this study, I assume almost all items have roughly equal difficulty. The reason is that all statements are mild; there is no strong statement that is able to distinguish top students or the poorest performance students very well. All items are expected to be agreed with by similar amount of students because I assume that students with higher than average levels of mathematics self-concept are highly possible to agree the majority items. Agreement/disagreement with those items do not require very high levels/low levels of mathematics self-concept. However, some slight differences are expected to exist among those item difficulties. The expected order as shown below:

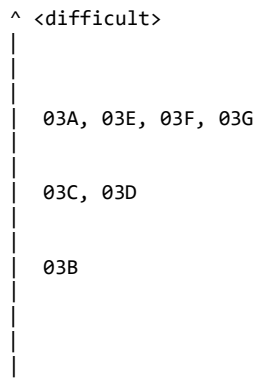

Item 03A, 03E, 03F, and 03G are possible to be harder to agree with comparing with other items. 03A (I usually do well in mathematics) is expected to be harder to agree with since that Taiwan's math tests are usually difficult, which makes students relatively hard to "usually do well in mathematics". Therefore, people with relatively high levels of mathematics self-concept can agree with 03A, and people with medium levels or low levels of mathematics self-concept may disagree with this item. 03E (I am good at working out difficult mathematics problems) is expected to be harder to agree with because difficult mathematics problems in Taiwan are possible more difficult than other places'. Only people with relatively high levels of mathematics self-concept can agree with this item. 03G (Mathematics is harder for me than any other subject) is a reversed item; it is assumed to be harder to disagree with because of the similar reason of 03A. Math is usually one of the most difficult subjects for Taiwan's students, which indicates that only people who have high levels of mathematics self-concept will disagree with this item. However, I don't expect the majority students will disagree with 03A and 03E or agree with 03G although Math tests in Taiwan are usually more difficult than other countries'. As stated above, students in Taiwan receive more intensive mathematical trainings, and are expected to have higher performance in math achievement tests and feel more confident about their mathematical ability. Therefore, I assume that students with relatively high levels of mathematics self-concept gain higher points on these items. 03F (My teacher tells me I am good at mathematics) is also possibly harder to agree with because teachers in Taiwan do not praise students as often as teachers in Western countries do. So I assume only students who have good performance in Math will be told by teachers that they are good at mathematics.

      03C and 03D are supposed to be easier to agree with comparing with 03A, 03E, 03G, and 03F. 03C (I'm just not good at mathematics) is a reversed item. It is not a strong statement that will be disagreed with by high levels of mathematics self-concept people only. I assume that people with around or above average levels of mathematics self-concept will deny 03C. Those people are not incapable enough to be counted as "not good at mathematics"; and the rest may agree with this item. 03D (I learn things quickly in mathematics) is approximately equal difficult to 03C. Nearly half of people learn things relatively quickly than the other 50% of people. So I also assume that people with above average levels of mathematics self-concept will agree with 03D, and people will disagree with this item if their mathematics self-concept is below the

average level.

03B (Mathematics is harder for me than for many of my classmates) is a reversed item. Only people with low levels of mathematics self-concept will agree with that math is harder for them than for many of their classmates. Therefore, 03B is relatively easy to disagree with.

In sum, people who agree with 03A, 03E, and 03F, and disagree with 03G are expected to agree with 03D, and disagree with 03C and 03B. Furthermore, agreement on 03D or disagreement on 03C indicates a disagreement about 03B.

In terms of the above explanation, a student who is good at mathematics should be able to usually do well in mathematics, good at working out difficulty mathematics problems, be told by teachers that he/she is good at mathematics, not feel math is harder than any other subject, deny that he/she is not good at mathematics, learn math things quickly, and not fell math is harder for himself/herself than many of classmates. On the contrary, a student who is not good at math should have opposite attitudes toward those statements.

There are some discrepancies between people who have high levels of mathematics self-concept and who have low levels of mathematics self-concept. According to PISA2012, students who have low levels of mathematics self-concept perform worse in mathematics than students who are more confident in their own abilities as mathematics learners. The relationship between students' self-concept and their mathematics performance was strong and positive. High levels of mathematics self-concept link more closely to good mathematics performance, while low levels of mathematics self-concept tend to relate to bad performance in math.

## Measurement model details

Rasch model is utilized in this study. It is represented by a logistic function:

$$P(X = 1| \beta, \delta) = \frac{e^{(\beta-\delta)}}{1+ e^{(\beta-\delta)}}$$, where β means person ability; δ means item difficulty. There

is no other information about the item or the person that is necessary for this model to estimate the probability of getting an item correct.

In this study, rating scale model is more appropriate to use given the fact that all the items here that exacted from TIMSS 2011 student questionnaire for Taiwan use the same response format: Strongly Disagree, Disagree, Agree, Strongly Agree. This means the test constructors, respondents, and test users all perceive those items to share the same rating scale. In this sense, rating scale model which is "one in which all items (or groups of items) share the same rating scale structure" (Linacre, 2000) is more reasonable than Partial Credit Model to be applied in this study. The equation of rating scale model is presented below:

$$\text{Rating Scale Model: } \pi_{nix} = \frac{e^{\sum_{j=0}^{x_{ni}}[\beta_n - (\delta_i + \tau_j)]}}{\sum_{k=0}^{m} e^{\sum_{j=0}^{x_{ni}}[\beta_n - (\delta_i + \tau_j)]}}$$

$\pi_{nix}$ represents the probability of person n responding in category x to item i, where $\beta_n$ is person n's ability; $\delta_i$ is the location or difficulty of item i on the variable; $\tau_j$ is the $k^{th}$ threshold location of the rating scale. These are constant across all items.

The primary assumptions of the Rasch model are local independence and unidimentionality. Local independence is an assumption in CTT, Rasch, and IRT. It requires that the items in a test should be independent of each other. To be specific, local independence means an item's correct or wrong answer should not be related to another item's correct or wrong answer under effects of any latent trait that is not the intended latent trait to be measured. Violation of local independence assumption will bring imprecise results. We will not be able to distinguish between the effects from target variable and which from unintended variable(s).

The other assumption of Rasch analysis is that the variable we intended to measure should be unidimensional. We use Rasch model to measure one thing only. If we have multidimension in Rasch analysis, it is very difficult to build a continuum because we have no idea about which continuum belongs to which dimension.

Rasch model can be considered as a simplified version of a 3-PL model. The equation for 3-PL model, which measures the probability of correctly answer a dichotomous item $i$, is:

$$P_{X=1}(\theta) = C_i + (1 + C_i)\frac{e^{a_i (\beta-\sigma_i)}}{1 + e^{a_i (\beta-\sigma_i)}}$$

where: $\beta$ is person ability; $\sigma_i$ is item $i$ difficulty; $a_i$ is the discrimination parameter; and $C_i$ is the guessing parameter. Rasch model does not consider the effects of guessing and discrimination; thus, we can remove guessing parameter and discrimination parameter from the 3-PL model and then obtain a Rasch model. Therefore, Rasch model can be considered as a simplified version of a 3-PL model.

- The principles of Rasch measurement instrument development and its purpose

According to the slides of Rasch General Principles, seven principles of Rasch measurement are listed below:

(a) the items should be unidimensional;

(b) the items should vary from very easy to very difficult;

(c) uniform spread of items along a continuum is required;

(d) the items should be hierarchical in the nature of their progression along the continuum;

(e) the items should be of equal discrimination;

(f) the items should be local independent in the sense that an answer to one is not dependent upon the answer to another; and

(g) "weeding" should be conducted so that "they on the whole fit well".

The purpose of Rasch measurement models and purposes of 2-PL and 3-PL models are significant different. In Rasch model, if a variable may be hypothesized to exist as a unidimensional construct, then it is possible to develop a hierarchical series of items that increase from a low level of difficulty (either easy to accomplish or endorse) to a higher level of difficulty (either harder to accomplish or endorse). The Rasch model is used as a criterion for the structure of the responses of those items. Whereas, The 2-PL and 3-PL models are used to seek to maximize the extent to which specific item response patterns can be reproduced, and to reduce residual variation. Therefore, 2-PL and 3-PL models can always "fit" any data set better than Rasch models.

Rasch model is considered at first by researchers because of its advantages that can provide particular insights. Rasch model provides an opportunity to order the items along a continuous scale that is invariant in terms of level of ease or difficulty or accomplishing the task for an individual being tested. In addition, Fox & Jones (1998) pointed out that "Rasch modeling allows for generalizability across samples and items, takes into account that response options may not be psychologically equally spaced, allows for testing of unidimensionality, produces an ordered set of items, and identifies poorly functioning items as well as unexpected responses". All of these benefits bring Rasch model to researchers at the first place.

- The expected value of items of each person

The expected value of a person on an item can be computed through the equation of rating scale model:

$$\text{Rating Scale Model: } \pi_{nix} = \frac{e^{\sum_{j=0}^{x_{ni}}[\beta_n - (\delta_i + \tau_j)]}}{\sum_{k=0}^{m} e^{\sum_{j=0}^{x_{ni}}[\beta_n - (\delta_i + \tau_j)]}}$$

When $\beta_n$ (person n's ability), $\delta_i$ (difficulty of item i on the variable), and $\tau_j$ (the k$^{th}$ threshold location of the rating scale) are known.

Persons and items with perfect correct or zero scores are removed from analysis because mathematical problems. In PROX initial estimation, estimation of person can be computed through the following model:

$$b_v = H + X(\ln \frac{r_v}{L-r_v}) = H + \sqrt{1 + \frac{\omega^2}{1.7^2}} \, (\ln \frac{r_v}{L-r_v})$$

where H is the mean of item difficulty; $\omega^2$ is item difficulty variance; L refers to number of items; and $r_v$ is person total score. In this model, perfect correct scores mean the person gets all items correct, indicating a full total score for L items. So the person total score is L *1 = L. Therefore, estimation of person $b_v = H + X(\ln \frac{r_v}{L-r_v}) = H + X(\ln \frac{L}{L-L}) = H + X(\ln \frac{L}{0})$.

Since 0 can never be a denominator, so persons with perfect correct scores will all be removed. Similarly, person with zero correct scores missed every single item, which means the person raw score is 0. Therefore, estimation of person who has zero correct scores $b_v = H + X(\ln \frac{r_v}{L-r_v})$

$= H + X(\ln \frac{0}{L-0}) = H + X(\ln(0))$. ln(0) equals negative infinity, which will bring estimation of person $b_v$ to be negative infinity as well. However, it is not acceptable to use negative infinity estimation of person. Therefore, we will remove persons with zero correct scores as well.

For estimation of item, items with perfect correct or zero scores will be removed as well. The estimation of item can be obtained through the following model:

$$d_i = M + Y[\ln (\frac{N-S_i}{S_i})] = M + \sqrt{1 + \frac{\sigma^2}{1.7^2}} \, [\ln (\frac{N-S_i}{S_i})],$$

where M is the mean of person ability; $\sigma^2$ is person ability variance; N refers to number of persons; and $S_i$ is item total score. In this model, perfect correct scores mean the item total score is N*1=N. Therefore, estimation of item $d_i = M + Y[\ln (\frac{N-S_i}{S_i})] = M + Y[\ln (\frac{N-N}{N})] = $

$M + Y[\ln(0)]$. As stated above, $\ln(0)$ equals to negative infinity, which will bring estimation of item $d_i$ to be negative infinity as well. So we will not keep perfect correct scores items. For zero correct score item, the estimation of item $d_i = M + Y[\ln(\frac{N-S_i}{S_i})] = M + Y[\ln(\frac{N-0}{0})] = M + Y(\ln\frac{0}{0})$. Since 0 can never be a denominator, so items with perfect correct scores will also be removed.

- "Sufficient Statistic"

    For a statistical model that conditioned on an unknown parameter $\theta$, a sufficient statistic is a function $T(X)$ that contains perfect information needed to compute any estimate of the parameter $\theta$. There is no other statistic that is able to provide any additional information to the value of the parameter $\theta$. Take UCON as an example. When we go through the operation, the original equation is developed to:

$$\Lambda = p\{X_{vi}|\beta_v, \delta_i\} = \frac{e^{\sum_v^N \sum_i^L X_{vi} \beta_v - \sum_v^N \sum_i^L X_{vi} \delta_i}}{\prod_v^N \prod_i^L (1+e^{\beta_v - \delta_i})}.$$

where
$$X_{vi} = \begin{cases} 1 \text{ if correct} \\ 0 \text{ if incorrect,} \end{cases}$$

$\beta_v$ = person ability parameter

$\delta_i$ = item difficulty parameter,

L = the number of items,

N = the number of persons with test scores between 0 and L.

Since $x_{vi}$ is person's score in each item, $\sum_i^L x_{vi} = r_v$, where $r_v$ is person total score. Similarly, for item estimation, $\sum_i^L X_{vi} = s_i$, where $s_i$ is item total score. Therefore, $r_v$ and $s_i$ are sufficient statistics for the maximum likelihood estimation matrix.

- Person and item "logits"

    Before explaining what person and item logits are, the concept of odds should be defined. According to course slides, odds means probability of event occurring over probability of event not occurring. In Rasch, odds turn to be the probability of getting items right over probability of getting items wrong. Logit, in general, refers to log odds: $\text{Log}_e[(\text{Probability of Success})/(\text{Probability of Failure})]$. Here is the equation:

$$\text{Logit} = \ln\left(\frac{P(\text{right})}{P(\text{wrong})}\right)$$

    Then a person logit refers to log odds of a person getting an item with a zero logit difficulty correct. Thus, the higher ability a person has, the higher the logit value. Similarly, an item logit means log odds of a person with a zero logit ability getting an item wrong. So the harder the item, the higher the logit value.

- Procedure of Maximum Likelihood Estimation operates.

    Maximum-likelihood estimation is a method of estimating the parameters of a particular model given data. There are two types of maximum likelihood estimation: unconditional maximum likelihood estimation (UCON) and conditional maximum likelihood estimation.

    In Rasch model for dichotomous items, UCON refers to the probability of a response pattern $X_{vi}$ to item i by person v. The likelihood of the matrix $X_{vi}$ is the continued product of individual person by item probabilities over all values of v and i. The procedure of operating maximum likelihood estimation is shown in the below:

[1]
$$p\{X_{vi}|\beta_v, \delta_i\} = \prod_v^N \prod_i^L \frac{e^{[X_{vi}(\beta_v - \delta_i)]}}{1+e^{\beta_v - \delta_i}} = \Lambda$$

where
$$X_{vi} = \begin{cases} 1 \text{ if correct} \\ 0 \text{ if incorrect,} \end{cases}$$

$\beta_v$ = *person ability parameter*

$\delta_i$ = *item difficulty parameter,*

L = the number of items,

N = the number of persons with test scores between 0 and L.

Then equation [1] can be expanded to:

[2]
$$\Lambda = p\{X_{vi}|\beta_v, \delta_i\} = \prod_v^N \prod_i^L \frac{e^{[X_{vi}(\beta_v - \delta_i)]}}{1 + e^{\beta_v - \delta_i}} = \frac{\prod_v^N \prod_i^L e^{[X_{vi}(\beta_v - \delta_i)]}}{\prod_v^N \prod_i^L (1 + e^{\beta_v - \delta_i})}.$$

Equation [2] can be replaced by exponentiation of a summation in the exponent:

[3]
$$\Lambda = p\{X_{vi}|\beta_v, \delta_i\} = \frac{\prod_v^N \prod_i^L e^{[X_{vi}(\beta_v - \delta_i)]}}{\prod_v^N \prod_i^L (1 + e^{\beta_v - \delta_i})} = \frac{e^{\sum_v^N \sum_i^L [X_{vi}(\beta_v - \delta_i)]}}{\prod_v^N \prod_i^L (1 + e^{\beta_v - \delta_i})}.$$

Through summation and multiplication, equation [3] is changed to:

[4]
$$\Lambda = p\{X_{vi}|\beta_v, \delta_i\} = \frac{e^{\sum_v^N \sum_i^L [X_{vi}(\beta_v - \delta_i)]}}{\prod_v^N \prod_i^L (1 + e^{\beta_v - \delta_i})} = \frac{e^{\sum_v^N \sum_i^L X_{vi}\beta_v - \sum_v^N \sum_i^L X_{vi}\delta_i}}{\prod_v^N \prod_i^L (1 + e^{\beta_v - \delta_i})}.$$

Let $\sum_v^N \sum_i^L x_{vi} \beta_v = \sum_v^N r_v \beta_v$, where $r_v$ *is person total score,*

$$\sum_v^N \sum_i^L X_{vi}\ \delta_i = \sum_i^L s_i \delta_i, \text{ where } s_i \text{ is item total score.}$$

Therefore, equation [4] is replaced by:

[5]
$$\Lambda = p\{X_{vi}|\beta_v, \delta_i\} = \frac{e^{\sum_v^N \sum_i^L X_{vi} \beta_v - \sum_v^N \sum_i^L X_{vi} \delta_i}}{\prod_v^N \prod_i^L (1+e^{\beta_v - \delta_i})} = \frac{e^{[\sum_v^N r_v \beta_v - \sum_i^L s_i \delta_i]}}{\prod_v^N \prod_i^L (1+e^{\beta_v - \delta_i})}$$

Based on $e^{x+y} = e^x * e^y$, equation [5] can be presented as following:

[6]
$$\Lambda = p\{X_{vi}|\beta_v, \delta_i\} = \frac{e^{[\sum_v^N r_v \beta_v]} * e^{[-\sum_i^L s_i \delta_i]}}{\prod_v^N \prod_i^L (1 + e^{\beta_v - \delta_i})}$$

Since that $\ln e^\beta = \beta$, therefore the log likelihood becomes:

[7]
$$\lambda = \log \Lambda = \ln \frac{e^{[\sum_v^N r_v \beta_v]} * e^{[-\sum_i^L s_i \delta_i]}}{\prod_v^N \prod_i^L (1 + e^{\beta_v - \delta_i})}$$

$$= \sum r_v \beta_v - \sum s_i \delta_i - \sum_i^L \sum_v^N (1 + e^{\beta_v - \delta_i})$$

Then we can obtain the first derivative and second derivative of λ with respect to $\beta$ and $\delta$. The first derivative of λ means the slop of the likelihood function for any given value of $\beta$ or

$\delta$. By setting the first derivatives of $\lambda$ equal to zero, we can then obtain maximum likelihood of the matrix $X_{vi}$. Particular value of $\beta$ & $\delta$ can be achieved through solving the equation when the first derivative of $\lambda$ equal to zero given the value of $\beta$ and $\delta$.

The first derivatives of $\lambda$ with respect to person ability $\beta$ is shown following:

[8]
$$\frac{d\lambda}{d\beta_v} = r_v - \sum_i^L p_{vi}$$

And the first derivatives of $\lambda$ with respect to item difficulty $\delta$ is:

$$\frac{dL}{d\delta} = -s_i + \sum_v^N p_{vi}$$

Where $$p_{vi} = \frac{e^{(\beta_v - \delta_i)}}{1 + e^{\beta_v - \delta_i}}.$$

For the first derivatives of $\lambda$ with respect to person ability $\beta$, $\sum_i^L p_{vi}$ means each person's estimated total raw score given their estimate of $\beta$. The reason of this interpretation is that the probability of getting a dichotomous item right is the expect score on that item. And for the first derivatives of $\lambda$ with respect to item difficulty $\delta$, $\sum_i^L p_{vi}$ means each item's estimated total raw score given their estimate of $\delta$.

The second derivatives ($d^2$) with respect to person ability $\beta$ becomes:

[9]
$$\frac{d^2\lambda}{d^2\beta_v} = -\sum_i^L p_{vi}(1 - p_{vi})$$

And the standard error for $\beta$ becomes:

$$SE_{\beta_v} = -[-\sum_i^L p_{vi}(1-p_{vi})]^{-\frac{1}{2}} = \frac{1}{\sqrt{\sum_i^L p_{vi}(1-p_{vi})}}$$

Similar to the equation of first derivative of $\lambda$ with respect to person ability $\beta$, $\sum_i^L p_{vi}$ means each person's estimated total raw score given their estimate of $\beta$.

The same situation happens in the second derivatives ($d^2$) with respect to item difficulty $\delta$:

$$\frac{d^2\lambda}{d^2\delta_i} = -\sum_i^L p_{vi}(1-p_{vi})$$

And the standard error for $\delta$ becomes the negative, inverse, square root of the second derivative:

$$SE_{\delta_i} = -[-\sum_r^{L-1} n_r p_{ri}(1-p_{ri})]^{-\frac{1}{2}} = \frac{1}{\sqrt{\sum_r^{L-1} n_r p_{ri}(1-p_{ri})}}$$

- Used PROX person ability and item difficulty estimates as initial estimation procedures

    PROX person ability and item difficulty estimates are used as initial estimation procedures because of its simplicity and efficiency. In PROX, number of items right for people and number of items right for items are the only thing needed to know when estimate person ability and item difficulty. They are the sufficient statistics to estimate a 1-parameter model. No other things are needed. The estimation equations are shown below:

$$d_i = \ln\frac{N-S_i}{S_i}$$

$$b_v = \ln\frac{r_v}{L - r_{v_i}}$$

where

$d_i$ = item estimate,

N = number of persons

$S_i$ = item total score = number of correct items

$b_v$ = person estimate,

L = number of items

$r_v$ = person total score = number of correct items

If we would like to put item and person estimates onto common scale in common units, the following equations can be helpful:

For estimation of person: $b_v = H + X(\ln\frac{r_v}{L-r_v}) = H + \sqrt{1 + \frac{\omega^2}{1.7^2}} \; (\ln\frac{r_v}{L-r_v})$,

where H is the mean of item difficulty; $\omega^2$ is item difficulty variance; L refers to number of items; and $r_v$ is person total score. This equation means: if a person has high abilities, the estimation of the person will be higher than the mean of item difficulty H, since the person is able to get more items right than wrong, thus $X(\ln\frac{r_v}{L-r_v})$ will be positive; however, if a person has low abilities, the estimation of the person will be lower than the mean of item difficulty H, since the person is able to get less items right than wrong so that $X(\ln\frac{r_v}{L-r_v})$ will be negative. The approximate standard error for person is also simple: $SE(\beta_V) = X\sqrt{\frac{L}{r_v(L-r_v)}} \approx \frac{2.5}{\sqrt{L}}$.

For estimation of item: $d_i = M + Y[\ln(\frac{N-S_i}{S_i})] = M + \sqrt{1 + \frac{\sigma^2}{1.7^2}} \; [\ln(\frac{N-S_i}{S_i})]$,

where M is the mean of person ability; $\sigma^2$ is person ability variance; N refers to number of persons; and $S_i$ is item total score. This equation means: a hard item will be above the mean of the people since it will lead to more people get it wrong, thus $Y \ln \left(\frac{N-S_i}{S_i}\right)$ will be positive; however, an easy item will be below the mean of the people since it will lead to less people get it wrong, thus $Y \ln \left(\frac{N-S_i}{S_i}\right)$ will be negative. The approximate standard error for item is also simple: $SE(d_i) = \frac{1}{\sqrt{\frac{N}{S_i(N-S_i)}}} \approx \frac{2.5}{\sqrt{N}}$.

Through the above method, we can quickly get a roughly precise initial value. Therefore, PROX estimates method is used to obtain initial estimation values.

- Procedure of the Newton-Raphson Algorithm

When the initial estimation value is obtained, the Newton-Raphson algorithm will then be used to make adjustments of the estimated value. New estimation can be computed by simply plugging the last estimation into the Newton-Raphson algorithm.

For persons' estimation, the equation is:

$$\hat{\beta}_{new} = \hat{\beta}_{last\ PROX} - \left[\frac{r - \sum_i^L p_{vi\ last}}{-\sum_i^L p_{vi\ last}(1 - p_{vi\ last})}\right]$$

where r means person raw score; $p_{vi\ last}$ is the 1st derivative; $p_{vi\ last}(1 - p_{vi\ last})$ is the 2nd derivative.

For items' estimation, the equation is:

$$\hat{\delta}_{new} = \hat{\delta}_{last\ PROX} - \left[\frac{(-S_i + \sum_v^{L-1} n_v p_{vi\ last})}{-\sum_v^{L-1} n_v p_{vi\ last}(1 - p_{vi\ last})}\right]$$

where $S_i$ means item raw score; $p_{vi\ last}$ is the 1st derivative; $p_{vi\ last}(1 - p_{vi\ last})$ is the 2nd derivative.

Through using the above equations, we will finally get a $\hat{\beta}_{\text{last PROX}}$ that roughly equals to $\hat{\beta}_{\text{new}}$, and a $\hat{\delta}_{\text{last PROX}}$ that roughly equals to $\hat{\delta}_{\text{new}}$, then the adjustment procedure is finished.

There is one problem that is worth to be mentioned. When using the Newton-Raphson algorithm to adjust the estimation, convergence is expected. It means the difference between two estimations should be decrease so that we will finally finish the adjustment. On the contrary, if divergence happens, the adjustment will never end, which is not what we expect.

- Person and item weighted fit statistics

Before actually measuring item/person, the data should be firstly examined to see if it can be brought into an order which fits with a measurement model. If not, we will not be able to use the data for measurement. INFIT, an information weighted sum, is one of the methods to examine the fit of data. INFIT looks at consistencies of responses over the entire set of items for each person in a person-level fit statistics and across the entire set of persons for each item for an item-level fit statistic. INFIT can be developed through the following procedure:

[1]

The expected value of any discrete variable is:

$$E(x) = \sum_{x=1}^{m} x \, f(x),$$

where x are the item values up to m possible;

f(x) is the probability of x occurring.

For the Rasch dichotomous responses of (0/1), the expected value is:

$$E(X_{vi}) = \sum_{x=0}^{1} x \, p_{X_{vi}} = 0 * p_0 + 1 * p_1 = p_1$$

$$E(X_{vi}) = p_{X=1}$$

For the polytomous scored items, the equation is identical:

$$E(X_{vi}) = \sum_{x=0}^{1} x\, p_{X_{vi}}$$

[2]

The variance of a discrete variable is:

$$V(x) = observed\ score - expected\ score = \sum_{x=1}^{m} [x - E(x)]^2 f(x)$$

For dichotomous items: $E(x) = P_{x=1}, f(X) = P_{x=x}$

Therefore, $V(x) = \sum_{x=0}^{1}(x - p_{x=1})^2 p_{x=x} = (0 - p_{x=1})^2 p_{x=0} + (1 - p_{x=1})^2 p_{x=1}$

Since $p_{x=0} = 1 - p_{x=1}$,

$$V(x) = (1 - p_{x=1})p_{x=1}^2 + (1 - p_{x=1})^2 p_{x=1} = p(1-p)$$

An identical process works for polytomous responses.

[3]

The standardized residual is:

$$Z_{vi} = \frac{raw\ residual}{standard\ deviation} = \frac{X_{vi} - P_{vi}}{\sqrt{V(x_{vi})}}$$

[4]

The INFIT equation bevomes:

$$V_i = \frac{\sum_{v=1}^{N} W_{vi} z_{vi}^2}{\sum_{v=1}^{N} W_{vi}}$$

$$W_{vi} = V(x_{vi}) = p(1-p)$$

There are various methods to interpret the INFIT results; interpretations mainly depend on investigators' experience. Usually, a value >1.4 indicates a relatively large number of unexpected responses have occurred. Also, the mean square can be transformed into an approximate t (ZSTD). An initial criterion of >2.0 is often used to highlight unusual response patterns. But it is also often changed to 3.0 or more depending on the sample size.

**Analysis**

- Variable Map

The variable that being measured is mathematics self-concept; it is students' beliefs in their own mathematics abilities. It is unidimensional and it varies along a hierarchical continuum. The seven items in this study measure a unidimensional latent variable, "mathematics self-concept". It has some similarities to mathematics self-efficacy. In terms of the definition (PISA, 2012), mathematics self-efficacy is also self-valuations about students' perceived ability to solve a range of mathematics problems. However, mathematics self-efficacy is used to describe students' beliefs about whether they can achieve certain academic achievement in the future through their efforts, while mathematics self-concept refers to students' belief in their ability to obtain certain levels of attainments in math after receiving mathematical education in school. In brief, mathematics self-concept focuses more on the outcome of learning; mathematics self-efficacy is more about the prediction of learning outcome.

Rasch models conceptualize items and people to vary along a hierarchical continuum. Each item provides qualitative information about what people are like at each position on the continuum. Items spread along continuum means that the items can differentiate across different levels of the continuum.

In this study, seven items indicate the extent to which a respondent's mathematics self-concept and they are expected vary along a hierarchical order. As previous section states, 03A, 03E, 03F, 03G are expected to have equal difficulty. From those four items to 03C and 03D, the degree of mathematical self-concept decreases. A person who agrees with 03A, 03E, 03F, 03G,

03C and 03D is supposed to have higher level of mathematical self-concept than a person who only agrees with the last two items 03C and 03D. What's more, a positive answer to the first four items will indicate (a) positive answer(s) to 03C and 03D. For example, an agreement to 03A "I usually do well in mathematics" indicates that the respondent will disagree with the reverse item 03C "I'm just not good at mathematics". In sum, responses to items that have different levels of difficulties represent respondent's different levels of mathematics self-concept. In addition, the respondents are ordered from lowest levels of mathematics self-concept to highest levels of mathematics self-concept based on the increasing scores.

In the variable map table 1.2 presented below, left-hand column shows that respondents are normally spread. People locate at the top of column receive high scores is those who have high levels of mathematics self-concept. They tend to agree with the statements of items or disagree with reversed directional items. Low scoring people who have low levels of mathematics self-concept are at the bottom of column. They would agree with fewer straightforward items or disagree with fewer reversed directional items. From the bottom to the top of right-hand column, the items tend to become more and more "difficult. For straightforward items, it would be difficult to agree with; and it would be difficult to disagree with if it was reversed direction. The items near bottom are most "easy" items to agree with or disagree with if it was reversed directional item. They are easy to endorse; while on the top of column are items that are most "difficult" to agree with/disagree with if it was reversed directional item, and hard to endorse. In table 1.2, 03G, a reverse directional item, is the hardest item to disagree with; while 03B, which is also a reverse directional item, is the easiest item to disagree with. They are identical to the hypothesized order. It is also reasonable to conclude that people at the top of column tend to agree with item all items or disagree with all reverse directional items because of their high levels of mathematics self-concept; people who are at the bottom of column are most likely to only disagree with 03G (reversed).

Moreover, as shown below, there is no big gap between seven items, which indicates that items have closed difficulties. This result is generally identical to our expectation that item difficulties of all items are close to each other.

```
TABLE 1.2 TIMSS2011 Student Questionnaire Math M ZOU209WS.TXT  Jan  4 2016 15:37
INPUT: 4026 PERSON  7 ITEM  REPORTED: 4026 PERSON  7 ITEM  4 CATS WINSTEPS 3.91.0
--------------------------------------------------------------------------------

MEASURE     PERSON - MAP - ITEM
                  <more>|<rare>
    1             .  +
                     |
                     |
                  .  |
                     |
                  .  |
                     |
                     |T
                .# T|
                     |
                .##  |   03GR(Math is harder for me than any other subject)
                     |S
            .####### S|   03A(I usually do well in math); 03CR(I'm just not good at math)
                     |
            .######## |   03F(My teacher tells me I'm good at math)
    0                +M
           .######### M|   03E(I'm good at working out difficult math problems)
                     |
         .############|   03D(I learn things quickly in math)
                     |S
              .###### S|
                     |
                 .##  |
                     |T 03BR(Math is harder for me than for many of my classmates)
                .# T|
                     |
                     |
                  .  |
                     |
                     |
   -1             .  +
                     |
                     |
                     |
                  .  |
                     |
                     |
                     |
                  .  |
                     |
                     |
                     |
   -2                +
                  <less>|<freq>
 EACH "#" IS 74: EACH "." IS 1 TO 73
```

Table 1.4 is the variable map of this study with scoring categories using a rating scale, and Table 3.2 contains summary of category structure. Through column of 50% CUM probability in Table 3.2, we can get estimates of threshold. The result of 50% CUM probability is slightly different from which in structure measure column, although they all provide estimation of the category threshold. The structure measures are Rasch-Andrich estimate for moving from 1 score to the next higher one and the measure is where there is a 50% probability of that transition happening. Whereas, the 50% CUM probability values are the Rasch-Thurstone estimate for moving from 50% probability in lower categories to higher ones, which is very useful for rating scale. Therefore, Rasch-Thurstone estimate is used to estimate threshold estimate of this study is that: -0.72 is the threshold from category 1 to 2; 0.01 is the threshold from category 2 to 3; and 0.72 is the threshold from category 3 to 4.

In variable map Table 1.4, the distribution of the persons and items is shown. The variable is laid out vertically with the highest scored persons (highest levels of Mathematics self-concept), and items that are most difficult to agree with at the top. Each item is shown three times in this table. In the center item column, each item is placed at its mean difficulty scale category. In the left-hand item column, the item is shown at the measure level (Rasch-Thurstone threshold) corresponding to a probability of .5 of being rated in the bottom rating scale category. In the right-hand item column, the item is shown at the measure level corresponding to a probability of .5 of being rated in the top rating scale category. In this study, most people score in the middle scale category level; some people locate on the bottom category level; while only few people on in the top scale category.

```
TABLE 3.2 TIMSS2011 Student Questionnaire Math M ZOU209WS.TXT  Jan  4 2016 15:37
INPUT: 4026 PERSON   7 ITEM  REPORTED: 4026 PERSON   7 ITEM   4 CATS WINSTEPS 3.91.0
--------------------------------------------------------------------------------

    SUMMARY OF CATEGORY STRUCTURE.  Model="R"
    ------------------------------------------------------------------
    |CATEGORY    OBSERVED|OBSVD SAMPLE|INFIT OUTFIT|| ANDRICH |CATEGORY|
    |LABEL SCORE COUNT %|AVRGE EXPECT| MNSQ  MNSQ||THRESHOLD| MEASURE|
    |------------------+------------+------------++---------+--------|
    |  1   1    7381  26|  -.26  -.24| 1.00  1.06||  NONE   |( -1.72)| 1
    |  2   2    7597  27|  -.11  -.12| 1.07  1.10||   -.21  |   -.45 | 2
    |  3   3    6962  25|   .02   .01|  .89   .86||    .03  |    .46 | 3
    |  4   4    6242  22|   .13   .15| 1.00   .99||    .19  |(  1.71)| 4
    ------------------------------------------------------------------
    OBSERVED AVERAGE is mean of measures in category. It is not a parameter estimate.

    ------------------------------------------------------------------------------
    |CATEGORY    STRUCTURE    |  SCORE-TO-MEASURE   | 50% CUM.| COHERENCE      |ESTIM|
    | LABEL    MEASURE  S.E.  | AT CAT. ----ZONE----|PROBABLTY| M->C C->M RMSR |DISCR|
    |-------------------------+---------------------+---------+----------------+-----|
    |  1       NONE           |( -1.72) -INF  -1.10 |         |  76%   1% 1.2907|     | 1
    |  2       -.21    .01    |   -.45 -1.10   .00  |   -.72  |  31%  63%  .5381| 1.17| 2
    |  3        .03    .01    |    .46   .00  1.10  |    .01  |  29%  51%  .5993| 1.01| 3
    |  4        .19    .02    |(  1.71) 1.10  +INF  |    .72  |  60%   1% 1.4104|  .83| 4
    ------------------------------------------------------------------------------
    M->C = Does Measure imply Category?
    C->M = Does Category imply Measure?

    --------------------------------------------------------------------------------
    | Category Matrix : Confusion Matrix : Matching Matrix                         |
    |         Predicted Scored-Category Frequency                                  |
    |Obs Cat Freq|      1          2          3          4   |    Total           |
    |------------+--------------------------------------------+--------------------|
    |          1 |   2429.35    2101.37    1613.06    1237.22 |   7381.00          |
    |          2 |   2105.58    2099.17    1836.22    1556.02 |   7597.00          |
    |          3 |   1607.56    1840.49    1815.34    1698.62 |   6962.00          |
    |          4 |   1236.84    1558.67    1699.43    1747.06 |   6242.00          |
    |------------+--------------------------------------------+--------------------|
    |    Total   |   7379.34    7599.70    6964.05    6238.91 |  28182.00          |
    --------------------------------------------------------------------------------
```

```
TABLE 1.4 TIMSS2011 Student Questionnaire Math M ZOU209WS.TXT  Jan  4 2016 15:37
INPUT: 4026 PERSON   7 ITEM  REPORTED: 4026 PERSON   7 ITEM  4 CATS WINSTEPS 3.91.0
--------------------------------------------------------------------------------

    MEASURE        | BOTTOM P=50% | MEASURE      | TOP P=50%   MEASURE
     <more> ----- PERSON -+- ITEM     -+- ITEM        -+- ITEM       <rare>
       2            +              +              +                    2
                    |              |              |
                    |              |              |
                    |              |              |
                    |              |              |
                    |              |              |
                    |              |              |
                    |              |              |
                    |              |              | 03GR
       1         .  +              +              +                    1
                    |              |              | 03A; 03CR
                 .  |              |              | 03F
                 .  |              |              | 03E
                    |              |              | 03D
                .#  |              |              |
               .##  |              | 03GR         |
           .###### |              | 03A; 03CR    | 03BR
          .######## |              | 03F          |
       0            +              +              +                    0
         .######### |              | 03E          |
      .############ |              | 03D          |
           .###### |              |              |
               .## | 03GR         |              |
                .# | 03A; 03CR    | 03BR         |
                   | 03F          |              |
                 . | 03E          |              |
                   | 03D          |              |
      -1        .  +              +              +                   -1
                   |              |              |
                 . | 03BR         |              |
                   |              |              |
                   |              |              |
                   |              |              |
                 . |              |              |
                   |              |              |
                   |              |              |
      -2            +              +              +                   -2
     <less> ----- PERSON -+- ITEM     -+- ITEM        -+- ITEM       <freq>
  EACH "#" IN THE PERSON COLUMN IS 74 PERSON: EACH "." IS 1 TO 73
```

Category characteristic curves is shown below in Table 21.1. As we estimated above, the thresholds for categories 1-2, 2-3, and 3-4, are -0.72, 0.01, and 0.72. So we can observe that when the value of people logits minus item logits is less than -0.72, people have high probability to score 1; people whose difference between person logits and item logits is located between 0.01 and -0.72 are more possible to score 2; people tend to score 3 when the difference between person logits and item logits is located on 0.72 to 0.01; and when the difference between person logits and item logits is greater than 0.72, people have high probability to score 4.

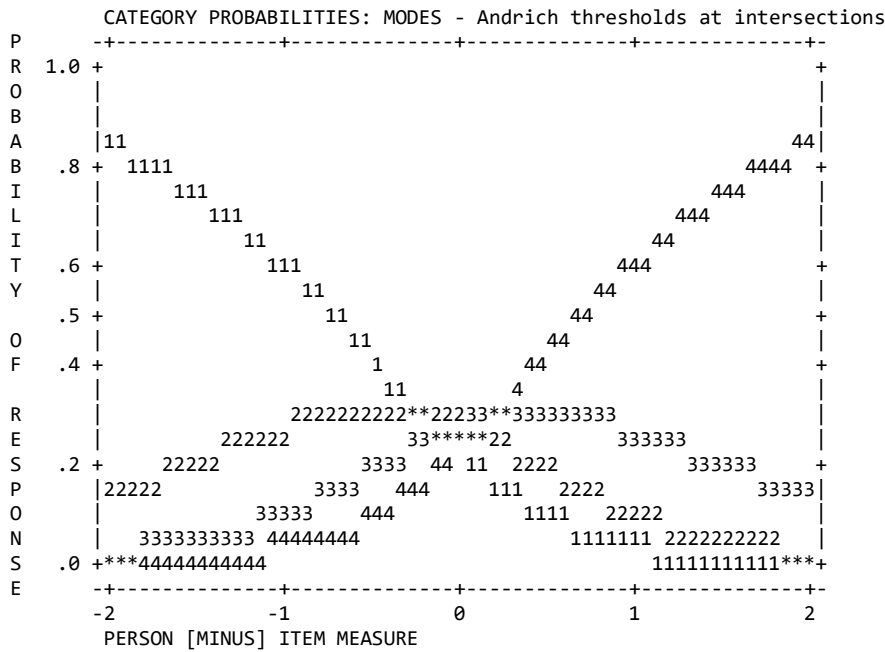

```
TABLE 21.1 TIMSS2011 Student Questionnaire Math   ZOU209WS.TXT  Jan  4 2016 15:37
INPUT: 4026 PERSON  7 ITEM  REPORTED: 4026 PERSON  7 ITEM  4 CATS WINSTEPS 3.91.0
-------------------------------------------------------------------------------
        CATEGORY PROBABILITIES: MODES - Andrich thresholds at intersections
P      -+--------------+--------------+--------------+--------------+-
R  1.0 +                                                              +
O      |                                                              |
B      |                                                              |
A      |11                                                          44|
B   .8 +  1111                                                   4444 +
I      |     111                                               444    |
L      |       111                                           444      |
I      |         11                                         44        |
T   .6 +          111                                     444         +
Y      |            11                                   44           |
    .5 +             11                                 44            +
O      |              11                               44             |
F   .4 +               1                              44              +
       |               11              4                              |
R      |             2222222222**22233**333333333                     |
E      |         222222          33*****22         333333             |
S   .2 +    22222             3333  44 11  2222         333333        +
P      |22222             3333    444    111  2222            33333|
O      |              33333    444         1111  22222               |
N      |   3333333333 44444444              1111111 2222222222       |
S   .0 +***44444444444                             11111111111***+
E      -+--------------+--------------+--------------+--------------+-
       -2             -1              0              1              2
         PERSON [MINUS] ITEM MEASURE
```

- The initial distributional characteristics of item difficulties and person abilities

In this study, the instrument is of medium "difficulty", which is corresponding to person "abilities". As shown in Table 1.2, left-hand column locates the person "ability" measures along the variable, and right-hand column locates the item "difficulty" measures along the variable.

Both distributions roughly cluster around logit 0. What's more, we can observe that the means of person distribution and item distribution are almost equal; one and two standard deviations of two distributions are also nearly of the same height. Therefore, it is reasonable to conclude that items and people seem to be on-target.

The findings are intended. As I clarified before, students in Taiwan are expected to have relatively higher levels of mathematics self-concept comparing with students in other countries. Hence, they should feel relative easy, at least not hard, to agree with these seven questionnaire items.

In this case, "difficult" item means items that hard to agree with and "easy" items means the statement of item are easy to be agreed with. Person "ability" defines person's mathematics self-concept: low "ability" people represent person with low levels of mathematics self-concept, and high "ability" people have high levels of mathematics self-concept.

- Misfit

It is crucial to understand that the Rasch model is a conceptual idea. No data is possible to fit the model perfectly. However, it is still necessary to check the degree to which the data is working with the model. If the data poorly cooperate with the model, we should consider reject the data because it will provide little information in our measurement. Fit statistics are indicators to guide us detect if there is any misfit problem. The source and reason of misfit data is worth to be further studied.

As Ludlow (2014) states, Rasch fit analyses mainly use residual statistics, checking the difference between a person's observed response on an item and the expected response under the model. A positive residual suggests a higher than expected response, while a negative residual indicate a lower than expected response. Fit discrepancy can be measured through mean-square residual statistics, such as OUTFIT and INFIT in Winsteps. OUTFIT is unweighted mean square, indicates outlier response over the entire set of items for each person and across the entire set of persons for each item. Weighted means square is so-called INFIT, looks at consistencies of responses over the entire set of items for each person in a person-level fit statistics and across the entire set of persons for each item for an item-level fit statistic.

There are various cutoffs suggested by various authors. In this study, the suggestions on Winsteps manual are adopted. Mean-squares show the size of the randomness. The expected values are 1.0. If values are less than 1.0, observations are probably too predictable; while values greater than 1.0 indicate unpredictability. Mean-squares usually average to 1.0, so if some are higher than 1.0, there should be also below ones. Zstd are used to test the hypotheses "do the data fit the model (perfectly)." 0.0 is the expected value. Similarly, less than 0.0 indicate too predictable, more than 0.0 indicates unpredictability. In general, mean-squares near 1.0 indicate little distortion of the measurement system, therefore, if the mean-squares values are fine, we don't need to look at Zstd. The manual puts forward that when mean-squares values are 0.5-1.5, the data is productive for measurement. By examining Table 10.1, we can see that both mean-squares values of INFIT and OUTFIT locate between 0.83 and 1.45, which indicate the data in this study probably do not suffer from misfit problems.

```
TABLE 10.1 TIMSS2011 Student Questionnaire Math  ZOU631WS.TXT  Jan  6 2016 11: 6
INPUT: 4026 PERSON   7 ITEM  REPORTED: 4026 PERSON   7 ITEM   4 CATS WINSTEPS 3.91.0
--------------------------------------------------------------------------------
PERSON: REAL SEP.: .00  REL.: .00 ... ITEM: REAL SEP.: 17.08  REL.: 1.00

            ITEM STATISTICS:  MISFIT ORDER

--------------------------------------------------------------------------------------
|ENTRY   TOTAL  TOTAL           MODEL|   INFIT  |  OUTFIT  |PTMEASUR-AL|EXACT MATCH|      |
|NUMBER  SCORE  COUNT  MEASURE  S.E. |MNSQ  ZSTD|MNSQ  ZSTD|CORR.  EXP.| OBS%  EXP%| ITEM |
|------------------------------------+----------+----------+-----------+-----------+------|
|    5   12003   4026    -.52    .02|1.39   9.9|1.45   9.9|A-.16   .26| 25.6  30.3| 03BR |
|    2   10678   4026    -.21    .02|1.03   2.0|1.04   2.7|B  .03  .27| 28.7  29.2| 03D  |
|    3   10152   4026    -.09    .02|1.02   1.5|1.03   1.9|C  .08  .27| 27.6  29.4| 03E  |
|    4    9371   4026     .09    .02| .98  -1.5| .98  -1.5|D  .63  .27| 24.1  29.8| 03F  |
|    6    8940   4026     .19    .02| .89  -7.2| .90  -6.0|c  .34  .27| 34.5  30.0| 03CR |
|    1    8930   4026     .20    .02| .86  -9.4| .86  -9.2|b  .57  .27| 31.1  30.0| 03A  |
|    7    8355   4026     .34    .02| .83  -9.9| .84  -9.2|a  .34  .26| 38.9  30.5| 03GR |
```

```
|----------------------------------+---------+---------+---------+---------+------|
| MEAN  9775.6 4026.0     .00    .02|1.00  -2.1|1.01  -1.6|          | 30.1  29.9|     |
| P.SD  1165.7     .0     .27    .00| .18   6.7| .19   6.5|          |  4.8   .4|     |
--------------------------------------------------------------------------------
```

Table 10.5 is also a very helpful table to diagnose misfit problems. Table 10.5 displays the item response strings with the most-unexpected responses. As the table shows, the majority of unexpected responses happen on reverse directional items: 03B, 03C, and 03G. High scoring students received low score in the "easiest" item 03B; while low scoring students had high score in "difficult" items 03C and 03G. This result suggests a possible problem of our method of putting straightforward items and reverse directional items into one scale. It also indicates a need to further discuss the reverse directional items.

```
TABLE 10.5 TIMSS2011 Student Questionnaire Math  ZOU631WS.TXT  Jan  6 2016 11: 6
INPUT: 4026 PERSON  7 ITEM  REPORTED: 4026 PERSON  7 ITEM  4 CATS WINSTEPS 3.91.0
--------------------------------------------------------------------------------

MOST UNEXPECTED RESPONSES
ITEM    MEASURE  |PERSON
                 |311 2111    3322111 3333333223333332111      32 1 3
                 |780 57548445491430797773207798653268749442104466 6
                 |58466532310865494828960501400136372960471517565  91
                 |74407645986307786183284565431839410995676410047925
              high------------------------------------------------
5 03BR     -.52 5|.11.11111111111111111111111....................
2 03D      -.21 2|1..11.1.............1.1........................
3 03E      -.09 3|.........1...11................................
4 03F       .09 4|................................44............4..
6 03CR      .19 6|............................44.4..444444..4..4..4
1 03A       .20 1|............................4................4....
7 03GR      .34 7|............................444444444444444.4...4.
                 |-----------------------------------------low
                 |31162111844332211173333333223333332111944213241931
                 |78005754310549143029777320779865326874471510466 65
                 |584 6532986865494888960501400136372960676417545 9
                 |744 7645    3077861 3284565431839410995     00 7 2
```
The reason of causing the majority of unexpected responses exist on reverse directional items is discussed as follows. As Hooper (2013) presents, "reverse directional items have different psychometric properties than the straightforward items" (p.1). Therefore, simply reverse coding the responses to these items cannot guarantee them to function as same as straightforward directional items do in measurement process.  Also, Hooper (2013)'s study indicates that the inclusion of reverse directional items may introduce construct irrelevant variance to the TIMSS context questionnaire scales, which may complicate the measurement model. Therefore, it is possible that many of the unexpected responses from fourth graders in Taiwan locate in 03B, 03C, and 03G.

- Dimensionality assumptions and parallel analysis

*Rasch principal component analysis (PCA) of residuals*

An ideal result of Rasch measurement is that all the data can be explained by the Rasch model. If so, the unexplained information that left over should be in random noise level. Rasch PCA of residuals is a way to look for patterns in the "unexpected" part of the data, which does not explained by the Rasch measurement. When conducting PCA of residuals, the hypothesis is that there is a common factor that explains the most residual variance. If this factor is discovered but only explains random noise, we can then conclude that there is no meaningful pattern in the residuals. Due to the divergent goals, principal-components analysis of residuals is not interpreted in the same way as common-factor analysis of original data does. The following section will present the results of my PCA of residuals.

In terms of Figure 1, the determinant is .016, which is what we expect because it indicates there is no linear dependence among the residuals. Also, KMO is .350, meaning the variance is not allowed to share and it is what we expect too. Furthermore, Bartlett's test of sphericity is 21059.206, which is significant to reject the null hypothesis and conclude that the correlation matrix is not an identity matrix.

*Figure 1*

**Correlation Matrix[a]**

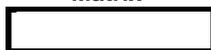

a. Determinant = .005

**KMO and Bartlett's Test**

| Kaiser-Meyer-Olkin Measure of Sampling Adequacy. | | .350 |
|---|---|---|
| Bartlett's Test of Sphericity | Approx. Chi-Square | 21059.206 |
| | df | 21 |
| | Sig. | .000 |

Next, we will take a look at Figure 2, the table of total variance explained, and Scree plot in Figure 3. As observing Figure 2, two components stand out with initial eigenvalues greater than 1. The 1$^{st}$ component is dominant by having eigenvalue of 3.920 and explaining 56.002% of variance. The 2$^{nd}$ one is relatively weak (eigenvalue = 1.341, % of variance = 19.156), but still stronger than the expectation. The anticipated average percentage of explained variance by each component is 14.286 given there are seven components. However, there are various criteria of threshold value; 1.00 is not always appropriate. For example, Smith and Miao (1994, p. 321) suggested 1.40 to be the cutoff for randomness. They observe many components with eigenvalues greater than one in four simulations of unidimensional observational data. In their simulations, the first component corresponds to the Rasch dimension. The eigenvalue of the second component, the largest component in the random noise, never exceeds 1.40. But Raîche G(2005) recommended to decide the criterion eigenvalue directly from relevant simulations because the value of 1.40 is always exceeded by the first eigenvalue, and usually by the second. Giving these previous studies, this study should have at least one shared component. The second one is remained questionable.

In my perspective, the recommendation of Raîche G(2005) is more reasonable. By checking the Scree plot (Figure 3), two dimensions apparently remains in the residual. What's more, according to the random data eigenvalues (Figure 4) which are computed using simulated data, the first component contains the largest eigenvalues. But the 95th percentile and mean random data eigenvalues of the first component are only 1.075330 and 1.058385 respectively. Thus, 1.40 is obviously too large to be the threshold in this study. Then therefore, first two components will be extracted in this case.

*Figure 2*

**Total Variance Explained**

| Component | Initial Eigenvalues | | | Extraction Sums of Squared Loadings | | |
|---|---|---|---|---|---|---|
| | Total | % of Variance | Cumulative % | Total | % of Variance | Cumulative % |
| 1 | 3.920 | 56.002 | 56.002 | 3.920 | 56.002 | 56.002 |
| 2 | 1.341 | 19.156 | 75.158 | 1.341 | 19.156 | 75.158 |
| 3 | .546 | 7.802 | 82.959 | | | |
| 4 | .443 | 6.327 | 89.286 | | | |
| 5 | .396 | 5.657 | 94.943 | | | |
| 6 | .321 | 4.586 | 99.529 | | | |
| 7 | .033 | .471 | 100.000 | | | |

**Total Variance Explained**

| Component | Rotation Sums of Squared Loadings | | |
|---|---|---|---|
| | Total | % of Variance | Cumulative % |
| 1 | 2.843 | 40.612 | 40.612 |
| 2 | 2.418 | 34.546 | 75.158 |
| 3 | | | |
| 4 | | | |
| 5 | | | |
| 6 | | | |
| 7 | | | |

*Figure 3*

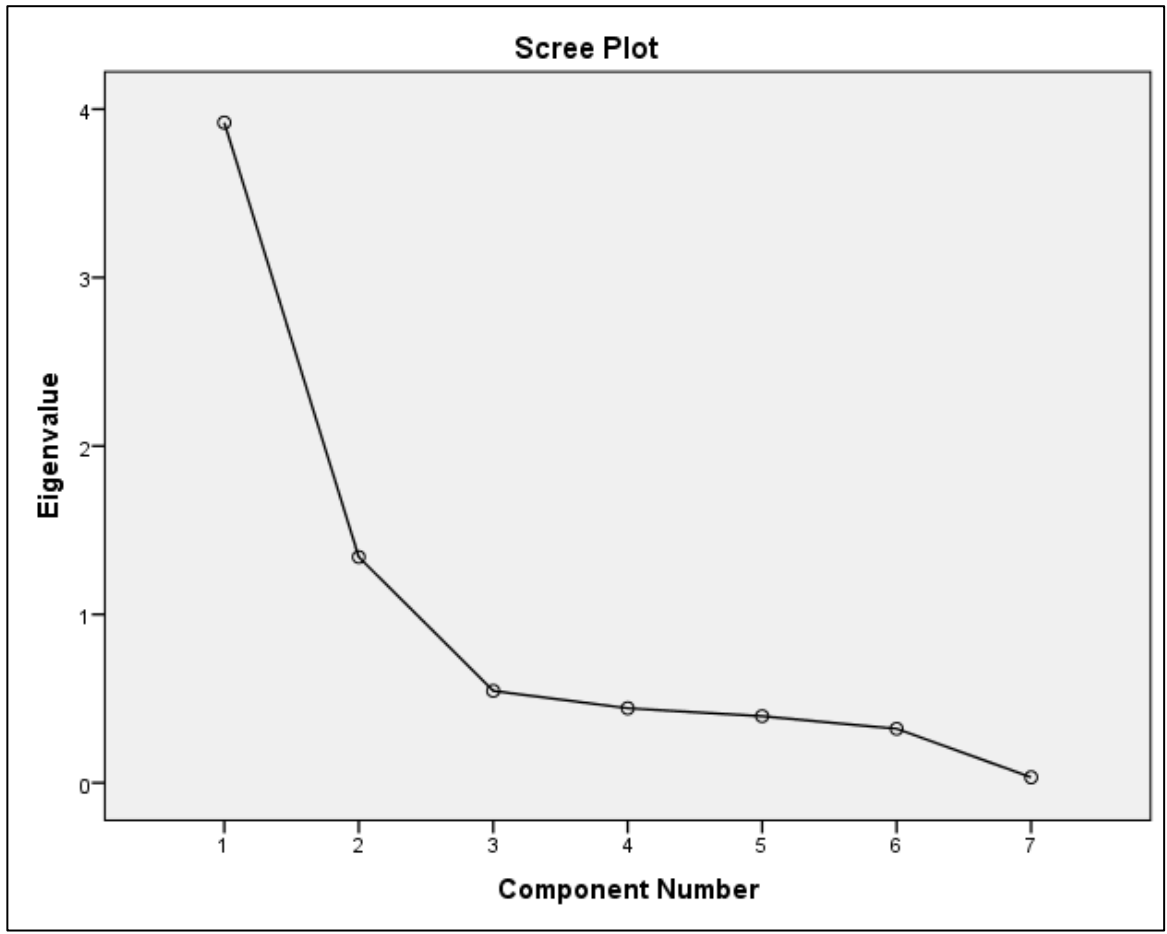

*Figure 4*

```
        Random Data Eigenvalues
    Root      Means     Prcntyle
 1.000000   1.058385   1.075330
 2.000000   1.035216   1.050564
 3.000000   1.017942   1.034192
 4.000000   1.000011   1.010672
 5.000000    .983497    .995187
 6.000000    .964006    .976827
 7.000000    .940944    .956422

      ------ END MATRIX -----
```

*Principal Component Parallel Analysis*

Parallel analysis can provide additional evidence about whether the residuals of Rasch measurement are random noise. In parallel analysis, the size of eigenvalues obtained from PCA is compared with random data eigenvalues, which obtained from a randomly generated data set of the same size. Only factors with eigenvalues exceeding the values obtained from the corresponding random data set are retained for further investigation.

In this study, the eigenvalues of simulated data is presented on Figure 4; and the results of principal component parallel analysis are shown from Figure 5 to Figure 8. According to the class PPT, a non-zero determinant, a small KMO, and a non-significant Bartlett test of sphericity are expected. As Figure 5 displays, the determinant of parallel analysis is .995 ≠ 0; KMO is .499, which is small; and the Bartlett's test of sphericity is non-significant with a p-value of 0.519. The results are exactly what we expect for the parallel analysis.

*Figure 5*

**Correlation Matrix**[a]

a. Determinant = .995

| KMO and Bartlett's Test | | |
|---|---|---|
| Kaiser-Meyer-Olkin Measure of Sampling Adequacy. | | .499 |
| Bartlett's Test of Sphericity | Approx. Chi-Square | 20.036 |
| | df | 21 |
| | Sig. | .519 |

Figure 6 shows eigenvalues that computed from the actual data, which are obtained through running a principal component analysis. By comparing with the eigenvalues computed from the random data sets (Figure 4), the eigenvalue of the first component (1.054) in Figure 6 is smaller than the corresponding first 95th percentile and mean random data eigenvalue (95th percentile: 1.058385; Mean: 1.075330); and the eigenvalue of the second component (1.039) is smaller than the corresponding first 95th percentile and slightly larger than mean random data eigenvalue (95th percentile: 1.035216; Mean: 1.050564). Therefore, we got evidence of no dominant $1^{st}$ component.

Additional evidences are in favor of random noise in residuals of our data. In terms of class PPT, the ratio of the $1^{st}$ to $2^{nd}$ eigenvalue is smaller than 3/1 together with the $1^{st}$ component accounts for less than 30% variance are two evidences of random noise in residuals. As Figure 6 showing, the eigenvalues of $1^{st}$ and $2^{nd}$ component are 1.054 and 1.039, and each account for 15.056% and 14.848% of variance. It is obvious that the eigenvalues and percentages of variance explained are roughly equally distributed across two components, which provides additional evidence of random noise.

*Figure 6*

**Total Variance Explained**

| Component | Rotation Sums of Squared Loadings | | |
| --- | --- | --- | --- |
| | Total | % of Variance | Cumulative % |
| 1 | 1.054 | 15.056 | 15.056 |
| 2 | 1.039 | 14.848 | 29.904 |

Extraction Method: Principal Component Analysis.

Figure 7 and 8 can also offer more evidence of random noise. A roughly circular pattern is displayed on Figure 7, the plot of the first two varimax rotated component loadings, which is exactly what we expect. Furthermore, no break is observed on scree plot of parallel analysis in Figure 8. Therefore, we get enough evidence and can be confidence to claim that the residuals of our Rasch measurement are random noise.

*Figure 7*

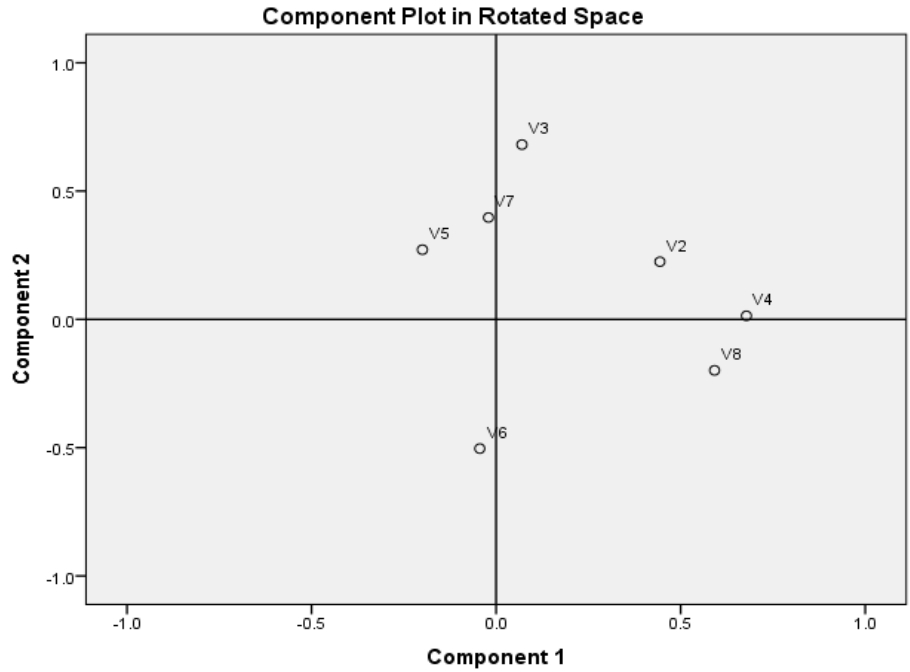

*Figure 8*

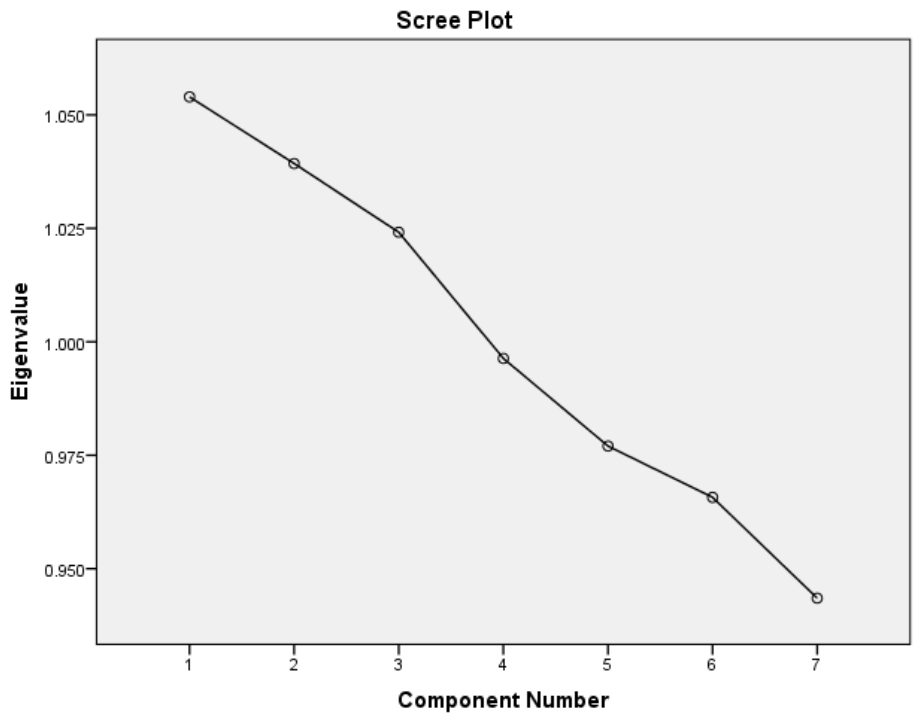

In conclusion, principal component parallel analysis provides evidence that the residuals are in random noise level, which perfectly meet our expectation. However, principal component analysis of residuals shows two shared components are extracted, indicating that the residuals are possible not as random as we expect. Therefore, the data in this study possibly fit the Rasch model well but have some risks.

## Conclusion

In general, the overall Rasch solution in this study is expected. As described on the first section, all items are expected to have little discrepancy between item difficulties, which is identical to the results in Table 1.2. All items spread approximately between 0.5 and -0.5. Moreover, distinctions between the hypothetical structure and actual results are reasonable. The hypothetical structure is:

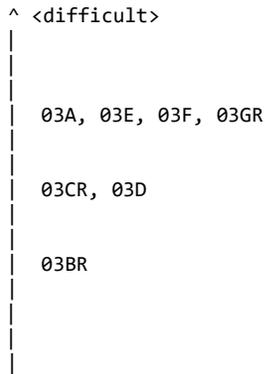

And the actual structure of the data is:

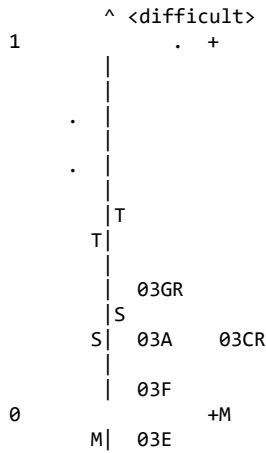

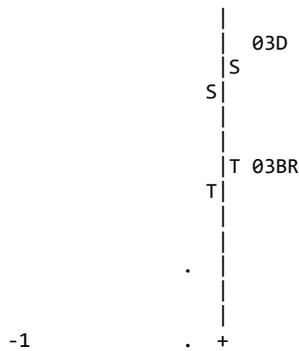

The order is a little bit different than the expectation. In hypothetical structure, 03A, 03E, 03F, 03GR are expected to have equal difficulty, which are slightly higher than 03C's and 03D's. Item difficulties of 03C and 03D are supposed to be equal too. As we see in Table 1.2, all items except 03A and 03C are displayed in a hierarchical order. However, their item difficulties are very closed to each other. The small discrepancy between item difficulties narrows the gap between the expected result and actual result. Therefore, the overall result is expected although some surprising results are showing up. And hence, I don't have recommendations of revising current instrument.

CTT results suggest that all items are easy to endorse (positive directional items are relatively easy to agree with and all negative directional items are easy to disagree with) according to item difficulties distribution. Also, person total score distribution shows that respondents tend to agree with items. Rasch results provide additional information about item difficulties and person ability. The distributions indicate that this instrument is of medium difficulty given the position of distributions along the continuum.

Rasch measurement results further offer information about poorly functioning items and unexpected responses, which are not explained by CTT results. In this Rasch analysis, little misfit problems are identified and random residuals are confirmed. However, the problem of reverse directional items containing most unexpected responses caught my eyes on the issues of reverse directional items. In the first article, straightforward items and reverse directional items are suggested to own different factors. The factor of straightforward items is "I'm good at Math; while the factor of reverse directional items is "I'm not good at Math". In Rasch analysis, reverse directional items are recoded and share the same measurement variable with straightforward items; the variable is mathematics self-concept. Although having only one factor for all items does not cause serious problems in this study, issues of analyzing reverse directional items are worth of further discussion. Is it precise to simply recoding the responses of reverse directional items when analyzing questionnaire scales? Will this method bring more problems when applied to other data?

# Appendix

**Reliability Statistics**

| Cronbach's Alpha | Cronbach's Alpha Based on Standardized Items | N of Items |
|---|---|---|
| .843 | .849 | 7 |

**Item-Total Statistics**

| | Scale Mean if Item Deleted | Scale Variance if Item Deleted | Corrected Item-Total Correlation | Squared Multiple Correlation | Cronbach's Alpha if Item Deleted |
|---|---|---|---|---|---|
| 03A_USUALLY DO WELL IN MATH | 13.70 | 23.267 | .685 | .543 | .810 |
| 03D_LEARN QUICKLY IN MATHEMATICS | 13.56 | 23.187 | .643 | .545 | .815 |
| 03E_GOOD AT WORKING OUT PROBLEMS | 13.43 | 23.141 | .652 | .569 | .814 |
| 03F_I AM GOOD AT MATHEMATICS | 13.30 | 23.281 | .624 | .521 | .817 |
| 03B_HARDER FOR ME THAN FOR OTHERS | 13.54 | 23.276 | .544 | .409 | .830 |
| 03C_JUST NOT GOOD IN MATH | 13.69 | 22.666 | .523 | .340 | .835 |
| 03G_MATHEMATICS HARDER FOR ME | 13.44 | 22.689 | .560 | .392 | .828 |

Factor Analysis:

**Correlation Matrix**[a]

a. Determinant = .050

**Total Variance Explained**

| Factor | Initial Eigenvalues | | | Extraction Sums of Squared Loadings | | | Rotation Sums of Squared Loadings[a] |
|---|---|---|---|---|---|---|---|
| | Total | % of Variance | Cumulative % | Total | % of Variance | Cumulative % | Total |
| 1 | 3.703 | 52.899 | 52.899 | 3.306 | 47.225 | 47.225 | 3.046 |
| 2 | 1.242 | 17.750 | 70.649 | .817 | 11.670 | 58.895 | 2.388 |
| 3 | .541 | 7.724 | 78.373 | | | | |
| 4 | .427 | 6.093 | 84.467 | | | | |
| 5 | .410 | 5.850 | 90.317 | | | | |
| 6 | .365 | 5.221 | 95.538 | | | | |
| 7 | .312 | 4.462 | 100.000 | | | | |

Extraction Method: Principal Axis Factoring.

a. When factors are correlated, sums of squared loadings cannot be added to obtain a total variance.

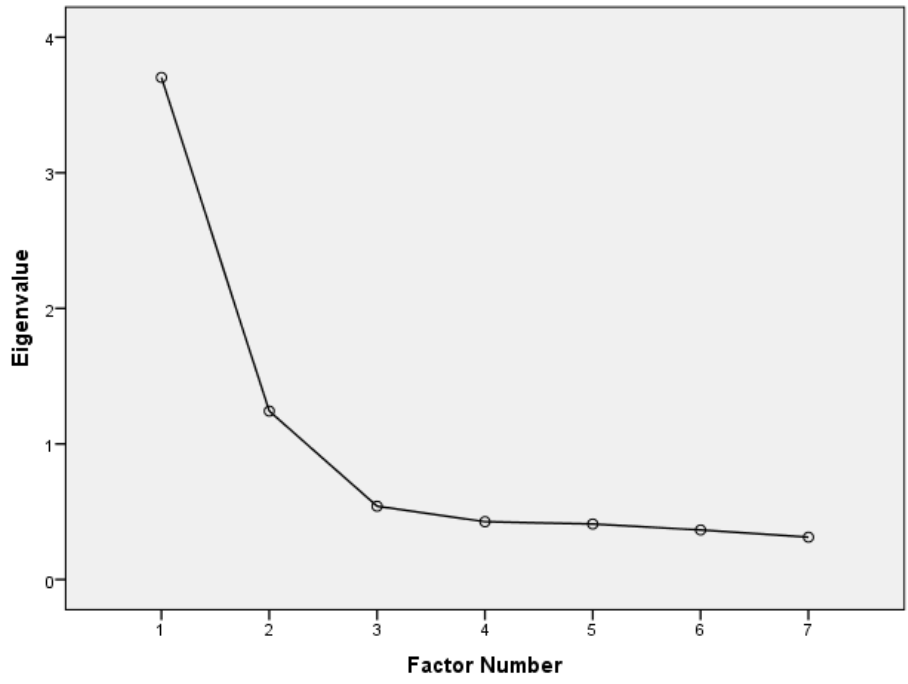